\documentclass[aps,prb,twocolumn,superscriptaddress]{revtex4-1}
\usepackage[hypertex]{hyperref}
\usepackage{amsmath}
\usepackage{amsfonts}
\usepackage{amssymb}
\usepackage{graphicx}
\usepackage{epsfig}
\usepackage{textcomp}
\usepackage{multirow}

\begin{document}

\title{Demonstrating Entanglement by Testing Bell's Theorem in Majorana Wires}
\author{David E. Drummond}
\affiliation{Department of Physics \& Astronomy, University of California, Riverside, California 92521, USA}
\author{Alexey A. Kovalev}
\affiliation{Department of Physics \& Astronomy and Nebraska Center for Materials and Nanoscience, University of Nebraska-Lincoln, Lincoln, NE 68588, USA}
\author{Chang-Yu Hou}
\affiliation{Department of Physics \& Astronomy, University of California, Riverside, California 92521, USA}
\affiliation{California Institute of Technology, Pasadena, California 91125, USA}
\author{Kirill Shtengel}
\affiliation{Department of Physics \& Astronomy, University of California, Riverside, California 92521, USA}
\author{Leonid P. Pryadko}
\affiliation{Department of Physics \& Astronomy, University of California, Riverside, California 92521, USA}
\date{\today}

\begin{abstract}
We propose an experiment that would establish the entanglement of Majorana zero modes in semiconductor nanowires by testing the Bell and Clauser-Horne-Shimony-Holt inequalities.  Our proposal is viable with realistic system parameters, simple ``keyboard'' gating, and projective measurement.  Simulation results indicate entanglement can be demonstrated with moderately accurate gate operations.  In addition to providing further evidence for the existence of the Majorana bound states, our proposal could be used as an experimental stepping stone to more complicated braiding experiments.
\end{abstract}
\maketitle

\section{Introduction}
The experimental observation of a self-conjugate fermionic particle has been a goal in physics since it was first theorized by Majorana over 75 years ago.\cite{Majorana:1937}  More recently, Majorana zero-energy modes bound to topological defects in 2D systems have emerged as candidates for a topologically protected qubit due to their non-Abelian braiding statistics,\cite{Read:2000, Ivanov2001,Nayak:Sep2008} and several groups have reported evidence suggesting that such Majorana bound states may exist at the ends of a semiconductor nanowire in the presence of $s$-wave superconductivity, a magnetic field, and strong spin-orbit coupling.\cite{Mourik:Science2012, Deng:NanoLett2012, Das:2012, Finck:Mar2013, Rokhinson:nov2012}  While this topological phase is theoretically supported by models and predicted in such systems,\cite{Kitaev:2001, Lutchyn:2010, Oreg:2010} further evidence is needed to rule out alternative explanations.\cite{Liu:Dec2012, sau:2012, Lee:Oct2012}

Perhaps the most definitive signature of Majorana bound states in these ``Majorana wires'' would be the demonstration of their non-trivial braiding statistics.  While braiding is ultimately needed for topological quantum computation, it remains an ambitious experimental task.  With this in mind, simpler experiments are desired to provide insight and direct further research before braiding is attempted.  Though there have been  feasible tests proposed and performed on several aspects of the system,\cite{Hassler:2010, Hassler:2011, Burnell:2013, DasSarma:2012, Rokhinson:nov2012, liu:PRB13, sau:2013} such as qubit measurement, there is still no clear consensus on the presence of Majorana bound states.\cite{franz:2013}  Observing entanglement of these states in Majorana wires would not only be a significant step towards their verification, but would also demonstrate their potential utility for topological quantum computation.  While tests of quantum entanglement  with Ising anyons (of which Majorana bound states are one example) have been discussed
formally,\cite{Brennen:2009, Campbell:2013} our goal is to devise and analyze a more concrete protocol motivated by recent experimental developments.

\begin{figure}[hbt]
 \begin{center}
   \includegraphics[width=\columnwidth]{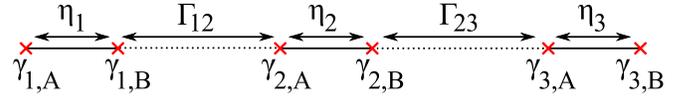}
 \caption{Gates are used to create three distinct topological regions (solid lines) in the wire.  Majorana bound states, represented by red x's, are localized at the ends of these regions.  Majorana bound states at the ends of the same topological region are coupled by $\eta$, while neighboring topological regions are coupled by $\Gamma$. }
   \label{fig:wire}
  \end{center}
\end{figure}

\begin{figure}[hbt]
 \begin{center}
   \includegraphics[width=0.93\columnwidth]{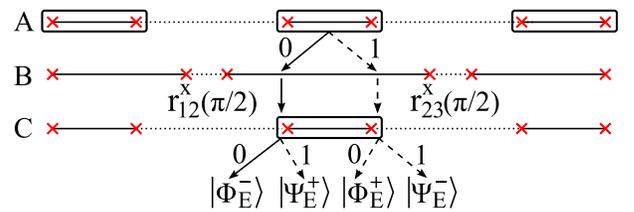}
 \caption{Preparation of the maximally entangled states of even total parity. A) The occupation of all three topological regions is measured, represented by rectangles around each region. B) The topological regions (solid lines) are expanded to perform $\pi/2$ rotations about the $x$-axis for the left and right logical qubits.  C) The middle qubit is measured, projecting to one of the four maximally entangled states of even total parity.  Different measurement outcomes are shown; when both middle measurements are $0$, the $|\Phi_\text{E}^{-}\rangle$ state is prepared as discussed in Sec.~\ref{sec:inequalities} (solid arrows), while other outcomes (dashed arrows) are discussed in Appendix \ref{sec:app2}.}
   \label{fig:bell-prep}
  \end{center}
\end{figure}

Thus we propose a procedure for demonstrating Bell's theorem with three pairs of Majorana bound states in semiconductor nanowire systems [see Fig.~\ref{fig:wire}].  Specifically, our procedure can be used to test the Bell\cite{bell:1964} and Clauser-Horne-Shimony-Holt (CHSH)\cite{clauser:69} inequalities using only two operations on maximally entangled states, which can be prepared using the same operations and a projective measurement [see Fig.~\ref{fig:bell-prep}].  These operations are accomplished by moving the domain walls along the axis of the wire using ``keyboard'' gates also needed for braiding operations.\cite{Alicea:Nat2011}  Hence, our proposal may also serve as a step towards experiments that perform topological operations. 

This paper will proceed as follows: in Sec.~\ref{sec:majorana} we introduce a simplified model for the Majorana wire and define a qubit basis. In Sec.~\ref{sec:inequalities} we summarize the entanglement inequalities and lay out the procedure for testing them. In Sec.~\ref{sec:simulation} we introduce a more realistic description of the semiconductor nanowire system, discuss corresponding simulation results, and introduce a simpler version of the CHSH experiment.  In Sec.~\ref{sec:experiment} we discuss experimental considerations and conclude in Sec.~\ref{sec:conclusion}.  We discuss modifications to our procedure for different measurement outcomes in Appendix \ref{sec:app2}, then review Bell's theorem and the entanglement inequalities in Appendix \ref{sec:app1}.

\section{Majorana Model Hamiltonian}
\label{sec:majorana}
To discuss the salient features of the Majorana wire system, we begin with a description similar to the toy model analyzed by Kitaev.\cite{Kitaev:2001}  With appropriate parameters, the wire is driven into a topological phase with an unpaired Majorana fermion at each end.\cite{Lutchyn:2010, Oreg:2010, Alicea:Nat2011}  If the parameters vary spatially (e.g., non-uniform chemical potential) there may be multiple topological regions separated by non-topological regions, with a Majorana fermion localized at each domain wall separating the two regions.  For our proposal, we will consider the case with three topological regions separated by two non-topological regions [see Fig.~\ref{fig:wire}] with the Majorana Hamiltonian 
\begin{multline}
H=i\eta_{1}\hat{\gamma}_{1,A}\hat{\gamma}_{1,B}+i\eta_{2}\hat{\gamma}_{2,A}\hat{\gamma}_{2,B}+i\eta_{3}\hat{\gamma}_{3,A}\hat{\gamma}_{3,B}\\
+i\Gamma_{12}\hat{\gamma}_{1,B}\hat{\gamma}_{2,A}+i\Gamma_{23}\hat{\gamma}_{2,B}\hat{\gamma}_{3,A} \label{eq:maj-ham},
\end{multline}
where $\eta$ describes the coupling between Majorana bound states at the ends of a single topological region, while $\Gamma$ describes the coupling of neighboring topological regions.  We assume that all couplings decay exponentially as the Majorana bound states separate from their nearest neighbors.  Each topological region has two types of Majorana operators, denoted by index $A$ or $B$, that form a conventional fermion operator $\hat{d}_{n}=\frac{1}{2}(\hat{\gamma}_{n,A}+i\hat{\gamma}_{n,B})$, and satisfy $\{\hat{\gamma}_{i},\hat{\gamma}_{j}\}=2\delta_{ij}$, where $i,j$ specifies both the region and type.  

The parity of the occupation number for the conventional fermions, (i.e., the eigenstate of $\hat{N}_{n}\equiv\hat{d}^{\dag}_{n}\hat{d}_{n}$), will serve as the degree of freedom for our qubits.  We specify a computational basis with the conventional fermions by defining the state $|000\rangle$ such that $\hat{d}_{n}|000\rangle=0$ for all $n$ and using the ordering conventions given by
\begin{eqnarray}
|000\rangle&\quad\quad\quad&|010\rangle=\hat{d}^{\dag}_{2}|000\rangle \notag \\
|011\rangle&=\hat{d}^{\dag}_{2}\hat{d}^{\dag}_{3}|000\rangle\quad\quad\quad&|001\rangle=\hat{d}^{\dag}_{3}|000\rangle\notag \\
|110\rangle&=\hat{d}^{\dag}_{1}\hat{d}^{\dag}_{2}|000\rangle\quad\quad\quad&|100\rangle=\hat{d}^{\dag}_{1}|000\rangle \label{basis} \\
|101\rangle&=\hat{d}^{\dag}_{1}\hat{d}^{\dag}_{3}|000\rangle\quad\quad\quad&|111\rangle=\hat{d}^{\dag}_{1}\hat{d}^{\dag}_{2}\hat{d}^{\dag}_{3}|000\rangle. \notag 
\end{eqnarray}
Since our model describes a system with superconductivity, the total number of particles is conserved modulo $2$.  This restriction splits the basis into two sub-bases, $S_{\text{E}}$ and $S_{\text{O}}$, with an even and odd number of total particles (i.e., total parity), which are the left and right columns of Eqs.~\eqref{basis} respectively.  A state from one basis cannot evolve into a state from the other basis since they differ by a single particle.  Strictly speaking, the two bases can interact if we account for quasi-particle poisoning\cite{Rainis:PRB13} in our model, but this occurs on a much longer time-scale than our proposed operations as discussed in Sec.~\ref{sec:experiment}.  We use the middle occupation number to preserve the total parity rather than storing unique quantum information.  Thus, two logical qubits are encoded in the left and right topological regions while the occupation of the middle region is forfetied as the ``parity qubit''.

By writing the Majorana operators in terms of the conventional fermions with $\hat{\gamma}_{n,A}=\hat{d}_{n}+\hat{d}^{\dag}_{n}$ and $i\hat{\gamma}_{n,B}=\hat{d}_{n}-\hat{d}^{\dag}_{n}$, the Hamiltonian in our basis is
\begin{multline}
H=-\eta_{1}(\sigma^{z}\otimes\sigma^{0}\otimes\sigma^{0})-\eta_{2}(\sigma^{0}\otimes\sigma^{z}\otimes\sigma^{0})-\eta_{3}(\sigma^{0}\otimes\sigma^{0}\otimes\sigma^{z})\\
-\Gamma_{12}(\sigma^{x}\otimes\sigma^{x}\otimes\sigma^{0})-\Gamma_{23}(\sigma^{0}\otimes\sigma^{x}\otimes\sigma^{x}).
\end{multline}
The $\eta$ terms for each topological region perform the $\sigma^{z}$ operation for their corresponding qubits, while the $\sigma^{x}$ operation is performed on the neighboring qubits involved in the $\Gamma$ terms.  Thus rotations of the qubits can be made by adjusting the parameters of the wire to suppress the couplings of all but one term in the Hamiltonian.  For example, if all the couplings other than $\Gamma_{12}$ are negligible, the evolution operator after time $T$ is
\begin{multline}
r^{x}_{12}(\theta)\equiv\exp{\left[i\frac{\theta}{2}(\sigma^{x}\otimes\sigma^{x}\otimes\sigma^{0})\right]}\\=\cos{\frac{\theta}{2}}(\sigma^{0}\otimes\sigma^{0}\otimes\sigma^{0})+i\sin{\frac{\theta}{2}}(\sigma^{x}\otimes\sigma^{x}\otimes\sigma^{0}),
\end{multline}
where $\theta=2\Gamma_{12}T/\hbar$ is the angle of the $XX$-rotation entangling qubits 1 and 2..  By adjusting the parameters appropriately, we can perform all the necessary operations for our proposal.

\section{Entanglement Inequalities}
\label{sec:inequalities}
Before testing the Bell and CHSH inequalities we discuss the preparation of one of the four maximally entangled states of even parity,
\begin{equation}
|\Phi_\text{E}^{\pm}\rangle=\frac{|000\rangle\pm|101\rangle}{\sqrt{2}},\quad\quad|\Psi_\text{E}^{\pm}\rangle=\frac{|011\rangle\pm|110\rangle}{\sqrt{2}}\label{eq:max-ent}
\end{equation}
using the operations already discussed and projective measurement.  To begin the preparation, the parity of each topological region is measured, fixing the total parity and projecting to one of the basis states.  The inequalities can be tested equivalently with any of the maximally entangled states from either parity, but for conciseness we assume the total parity is even for the rest of the body of this paper, and we only discuss the inequalities with $|\Phi_\text{E}^{-}\rangle$, assuming the initially measured state is $|000\rangle$.  Our proposal can be accomplished for general initial conditions by altering our procedure slightly as described in Appendix~\ref{sec:app2}.  If a $\pi/2$ rotation about the $x$-axis is performed for both logical qubits the resulting state is $r^{x}_{12}(\pi/2)r^{x}_{23}(\pi/2)|000\rangle$, or
\begin{equation}
\frac{|000\rangle-|101\rangle+i|011\rangle+i|110\rangle}{2}=\frac{|\Phi_\text{E}^{-}\rangle+i|\Psi_\text{E}^{+}\rangle}{\sqrt{2}},
\end{equation}
which will project to $|\Phi_\text{E}^{-}\rangle$ if the middle parity qubit is measured to be $0$.  Note that the $r_{12}^{x}$ and $r_{23}^{x}$ operations commute since they involve different $\gamma$ operators, making the operation order irrelevant (as well as allowing simultaneous operations).  In general, each maximally entangled state can be prepared by measuring all three qubits to project to a single basis state, extending the outer topological regions towards the middle topological region for a small time, returning them to their original position, then projectively measuring the middle qubit.

Once the state $|\Phi_\text{E}^{-}\rangle$ is prepared, we can test the version of Bell's inequality given in Appendix~\ref{sec:app1}, 
\begin{equation}
P_{=}(a,b)+P_{=}(b,c)+P_{=}(a,c)\geq1,\label{eq:bell}
\end{equation}
where $P_{=}(L,R)$ is the probability that the left and right qubits are equal after being rotated by angles $L$ and $R$, respectively.  The left side of the inequality, which we call the ``Bell quantity'', can be interpreted as the probability that at least one of the rotation combinations will make the left and right qubits equal. 

\begin{figure}[bht]
 \begin{center}
   \includegraphics[width=\columnwidth]{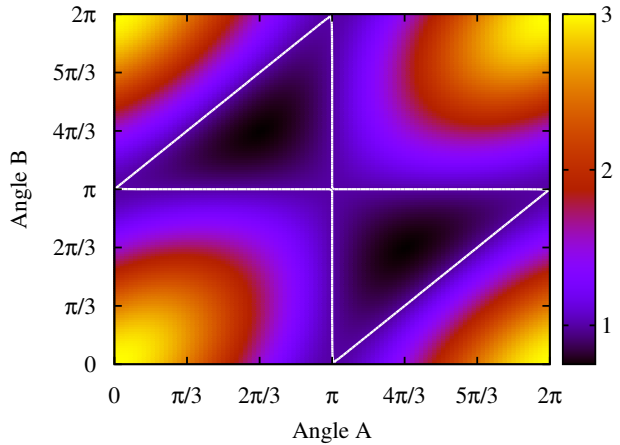}
 \caption{\label{fig:bell} Contour plot for the quantum mechanical prediction of the Bell quantity for the state $|\Phi_\text{E}^{-}\rangle$ defined in Eq.~\eqref{eq:max-ent}.  Local hidden variable theories require that the Bell quantity be greater than or equal to $1$, but it is predicted to be less than $1$ for relative rotation angles inside the white triangles. }
  \end{center}
\end{figure}

According to quantum mechanics the probability that the qubits are equal after rotations $L$ and $R$ is $\cos^{2}\left(\frac{L-R}{2}\right)$.  Only the relative angles between rotations are physically relevant, so we can set $A\equiv a-c$ and $B\equiv b-c$ to write the Bell quantity as  
\begin{equation}
\cos^{2}\left(\frac{A-B}{2}\right)+\cos^{2}\left(\frac{A}{2}\right)+\cos^{2}\left(\frac{B}{2}\right), \label{eq:bell-quant}
\end{equation}
which is plotted in Fig.~\ref{fig:bell}.  Quantum mechanics predicts the Bell quantity can be as low as $3/4$ (for the relative angles $A=2\pi/3$ and $B=4\pi/3$, or vice-versa) and is inconsistent with local hidden variable theories, which require the Bell quantity to be greater than or equal to $1$.  In principle, Bell's inequality could be experimentally tested in our proposal by repeatedly preparing maximally entangled states, performing the three rotation combinations in Eq.~\eqref{eq:bell}, and measuring the qubits to find the probability of each state.  

In practice however, almost every experiment that tests Bell's theorem uses the CHSH inequality discussed in Appendix~\ref{sec:app1},
\begin{equation}
\left|\langle L_{1},R_{1}\rangle+\langle L_{2},R_{1} \rangle+\langle L_{1},R_{2} \rangle-\langle L_{2},R_{2} \rangle\right|\leq2,\label{eq:chsh-orig}
\end{equation}
where $\langle L,R \rangle=P_{=}(L,R)-P_{\neq}(L,R)$ is the expectation value of the combined parity of the left and right qubits after being rotated by angles $L$ and $R$, respectively.  The left side of the inequality, which we call the ``CHSH quantity'', must be less than or equal to $2$ in local hidden variable theories.

\begin{figure}[bht]
 \begin{center}
   \includegraphics[width=\columnwidth]{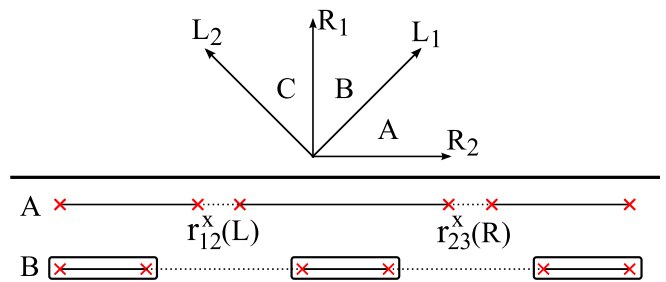}
 \caption{Top: Angles of rotation in CHSH inequality.  The left qubit is rotated by either angle $L_{1}$ or $L_{2}$, while the right qubit is rotated by either angle $R_{1}$ or $R_{2}$.  Bottom: A) One of the four rotation combinations is performed by extending the outer topological regions,  B) then returned for measurement. }
   \label{fig:angles}
  \end{center}
\end{figure}

According to quantum mechanics, the expectation value $\langle L,R \rangle=\cos(L-R)$ for general rotation angles $L$ and $R$ acting on $|\Phi_\text{E}^{-}\rangle$.  Again, only the relative angles of rotation are physically significant, so we introduce  angles $A\equiv L_{1}-R_{2}$, $B\equiv R_{1}-L_{1}$,  and $C\equiv L_{2}-R_{1}$, [see Fig.~\ref{fig:angles}], making the CHSH quantity
\begin{equation}
\left|\cos(A)+\cos(B)+\cos(C)-\cos(A+B+C)\right|,\label{eq:chsh-angle}
\end{equation}
which has a maximum of $2\sqrt{2}$ when $A=B=C=\pi/4$, contradicting the local hidden variable prediction.  The inequality can be tested experimentally by repeatedly preparing the state $|\Phi_\text{E}^{-}\rangle$, extending the topological regions to perform one of the four rotation combinations involved in Eq.~\eqref{eq:chsh-orig}, then returning the topological regions to their original position to measure the qubits.  

\section{Semiconductor Hamiltonian and Simulation}
\label{sec:simulation}
We now consider a more realistic model of the semiconductor system by introducing a one-dimensional lattice Hamiltonian
\begin{equation}
H_{\text{S}}=H_{\text{TB}}+H_{\text{SO}}+H_{\text{Z}}+H_{\text{SC}}, \label{ham-begin}
\end{equation}
where
\begin{eqnarray}
H_{\text{TB}}&=&\sum_{j\sigma}\left[(2t_{0}-\mu_{j})\hat{c}^{\dag}_{j\sigma}\hat{c}_{j\sigma}-t_{0}\hat{c}^{\dag}_{j\pm1,\sigma}\hat{c}_{j\sigma}\right],\\
H_{\text{SO}}&=&\sum_{j\sigma}\left[\alpha \,s(\sigma)\left(\hat{c}_{j\bar{\sigma}}^{\dag}\hat{c}_{j+1,\sigma}-\hat{c}_{j+1,\bar{\sigma}}^{\dag}\hat{c}_{j\sigma}\right)\right],\\
H_{\text{Z}}&=&\sum_{j\sigma}\left[s(\sigma) V^{z}\hat{c}^{\dag}_{j\sigma}\hat{c}_{j\sigma}+V^{s(\sigma)} \hat{c}_{j\bar{\sigma}}^{\dag}\hat{c}_{j\sigma}\right],\\
H_{\text{SC}}&=&\sum_{j}\left(\Delta \hat{c}_{j\uparrow}^{\dag}\hat{c}^{\dag}_{j\downarrow}+\Delta^{*}\hat{c}_{j\downarrow}\hat{c}_{j\uparrow}\right), \label{ham-end}
\end{eqnarray}
are the tight-binding, spin-orbit, Zeeman, and proximity-effect superconducting terms, respectively.  Here $\hat{c}_{j\sigma}$ is the electron annihilation operator for spin $\sigma$ at site $j$, $t_{0}=\hbar^{2}/(2m^{*}a^{2})$ is the tight-binding coefficient with effective mass $m^{*}$ and lattice size $a$, $\mu_{j}$ is the chemical potential at site $j$, $\alpha$ is the Rashba coupling, $\mathbf{V}=g\mu_{\text{B}}\mathbf{B}/2$ is the Zeeman coupling with $V^{\pm}=V^{x}\pm iV^{y}$ used for the terms perpendicular to the wire axis, and $\Delta$ is the $s$-wave pairing potential.  The coefficient $s(\sigma)$ stands for $+ $ and $-$ when $\sigma$ is $\uparrow$ and $\downarrow$, respectively, and $\bar{\sigma}$ denotes the opposite spin.

As shown by others,\cite{Lutchyn:2010, Oreg:2010, Alicea:Nat2011} this system has two topologically distinct phases.  When $\mu>\mu_{\text{T}}\equiv\sqrt{V_{\perp}^{2}-\Delta^{2}}$, where $V^{2}_{\perp}=(V^{z})^{2}+(V^{x})^{2}$ is the Zeeman field perpendicular to the spin-orbit quantization axis, the wire is a normal superconductor.  In the other case, $\mu<\mu_{\text{T}}$, a topologically distinct state emerges with Majorana bound states localized at the ends of the wire.  If the chemical potential varies spatially and crosses the topological limit at multiple locations, then multiple Majorana bound states will be present and the setup discussed in Sec.~\ref{sec:majorana} is possible.  

\begin{figure}[bht]
 \begin{center}
   \includegraphics[width=\columnwidth]{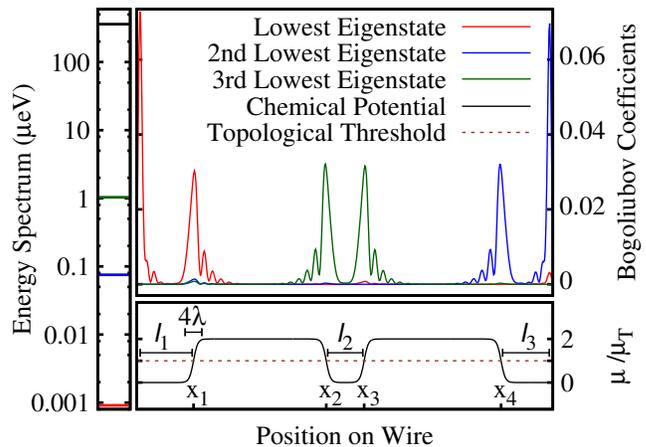}
 \caption{Bottom Right: A spatially varying chemical potential with three regions below the topological threshold of $\sqrt{V_{\perp}^{2}-\Delta^{2}}$, with domain wall lengths of $\sim4\lambda$.  Top Right: This leads to six Majorana bound states, one at the end of each region, that form three conventional eigenstates.  We plot the simulated spatial distribution of the Bogoliubov coefficients $\sum_{\sigma}(|u_{\sigma}|^{2}+|v_{\sigma}|^{2})$ along the length of the wire.  Left: The energy spectrum of these eigenstates is plotted in log scale, as well as the lowest-energy bulk state separated by a topological gap of $360 \mu$eV.  The splitting of the ``zero''-energy states is due to the exponentially small overlap in peaks, which is larger for the shorter topological regions.  There are also small disconnected peaks due to the basis choice of the simulation, which are mitigated by using slightly different lengths of $13.6\%$, $9.1\%$, and $12.6\%$ of the wire length for the left, middle, and right topological regions, respectively.}
   \label{fig:maj}
  \end{center}
\end{figure}

Specifically, our proposal separates the wire into three topological regions, leading to six Majorana bound states, one at the end of each region.  The Majorana bound states from each topological region can be paired together to form conventional fermions [e.g., $\hat{d}_{n}=(\hat{\gamma}_{A}+i\hat{\gamma}_{B})/2$] that correspond to three zero-energy (in the limit of an infinite wire) Bogoliubov excitations, separated from the higher-energy bulk states by a topological gap $\Delta_{\text{T}}$[see Fig.~\ref{fig:maj}].  Alternatively, these excitations can be thought of as the zero-energy eigenstate solutions to the Bogoliubov-de Gennes equations for the system.  Just as in the simpler model, the occupation number of the eigenstates localized to the left and right serve as the logical qubits, while the occupation number of the middle eigenstate does not contain unique quantum information since the total parity is conserved.

The spatial distribution of these excitations is contained in the coefficients $u$ and $v$ from the Bogoliubov transformation 
\begin{equation}
\hat{d}^{\dag}_{n}=\sum_{j\sigma}(u^{n}_{j\sigma}\hat{c}^{\dag}_{j\sigma}+v^{n}_{j\sigma}\hat{c}_{j\sigma}).\label{eq:bog}
\end{equation}
We begin our simulation by finding the coefficients for the lowest three eigenstates of the Hamiltonian in Eqs.~\eqref{ham-begin}-\eqref{ham-end} with parameters corresponding to  $m^{*}=0.015m_{\text{e}}$, $a=15$nm leading to $t_{0}=11.3$meV, $g=50$, $B=B^{z}=0.625$T leading to $V_{\perp}=0.9$meV, $\alpha=1.5$meV, and $\Delta=0.5$meV.  Thus the chemical potential marking the threshold between topological phases is $\mu_{\text{T}}=1.06$meV. The wire has $600$ sites leading to length $l=9\mu$m, with non-periodic boundary conditions.  At the domain walls the chemical potential smoothly alternates between $0$ and $2\mu_{\text{T}}$ over a length of approximately $4\lambda=0.04l$ with the profile function $\pm\mu_{\text{T}}\tanh(x/\lambda)
$, as shown in Fig.~\ref{fig:maj}.

Each of the lowest three eigenstates has two peaks localized at the ends of the topological region, indicating the location of the Majorana bound states.  Though the peaks decrease exponentially, their small, but non-zero overlaps cause the eigenstates to split from zero-energy.  Thus the  topological regions must be long enough to prevent these overlaps from splitting the excitations and coupling them to the bulk states.  

In addition to the Majorana bound state peaks at the domain walls of the expected topological region, our simulated eigenstates also have smaller, disconnected peaks at the other domain walls.  For example, the lowest eigenstate, which is mostly localized to the left, has a disconnected peak at the right end of the wire [see Fig.~\ref{fig:maj}].  These disconnected peaks are due to the implicit basis choice in our simulation rather than any physical phenomena.  

This can be understood by considering the symmetric case with periodic boundary conditions where all three topological regions have the same length, yielding a three-fold degeneracy.  For our proposal, the most useful basis is the set with each eigenstate localized in a single topological region.  However any normalized linear combination of the fully localized eigenstates is also an eigenstate, so the set of solutions found by our simulation may not be fully localized.

This basis ambiguity can be resolved by introducing asymmetry with slightly different topological region lengths.  As the topological region length shortens, the corresponding energy of the eigenstate localized to that region increases, splitting the degeneracy and localizing the simulated eigenstates.  Specifically we set the lengths between the domain walls as $l_{1}=x_{1}=0.136l$, $l_{2}=x_{3}-x_{2}=0.091l$, and $l_{3}=l-x_{4}=0.126l$.  We fine-tuned these lengths to split the degeneracy enough to ensure the eigenstates are nearly localized, yet not so much that the topological gap required for adiabatic dynamics is reduced too much.  Thus a few small disconnected peaks remain in our simulated eigenstates and the conventional fermions $\hat{d}_{n}$ in Eq.~\eqref{eq:bog} differ from the ideal fermions in Sec.~\ref{sec:majorana}.

One way to account for this difference is to consider the effect of additional couplings in the Majorana Hamiltonian in Eq.~\eqref{eq:maj-ham} while still using the ideal, fully localized operators $\hat{d}_{n}$.  For example, the lowest eigenstate's disconnected peak on the right end of the wire can be interpreted as effectively introducing several couplings, one of which is
\begin{equation}
i\eta_{13}\hat{\gamma}_{1,B}\hat{\gamma}_{3,B}=-\eta_{13}(\sigma^{x}\otimes\sigma^{z}\otimes\sigma^{y})
\end{equation}
where $\eta_{13}$ is a small coupling proportional to the disconnected peak size and we used the basis given in Eqs.~\eqref{basis} for the right hand side of the equation.  With this additional term, the operation $r_{12}^{x}$ becomes
\begin{eqnarray}
\tilde{r}_{12}^{x}&=&\exp\Big\{i\frac{t}{\hbar}\big[\Gamma_{12}(\sigma^{x}\otimes\sigma^{x}\otimes\sigma^{0}) \\
&&\quad\quad\quad\quad\quad\quad\quad-\eta_{13}(\sigma^{x}\otimes\sigma^{z}\otimes\sigma^{y})\big]\Big\}\notag \\
&=&r^{x}_{12}(\theta_{12})r_{1}^{x}(\theta_{13})r_{2}^{z}(\theta_{13})r_{3}^{y}(\theta_{13})r_{2}^{y}(\theta_{13}') \notag
\end{eqnarray}
where $r^{n}_{i}$ is a rotation about the $\hat{n}$-axis of the $i$th qubit's Bloch sphere, we define the angles $\theta_{12}=2\Gamma_{12}t/\hbar$, $\theta_{13}=-2\eta_{13}t/\hbar$, and $\theta_{13}'=2\eta_{13}\Gamma_{12}t^{2}/\hbar^{2}$, and we used the operator identity $\exp(A+B)=\exp(A)\exp(B)\exp(-[A,B]/2)$.  Euler's Rotation Theorem states that these rotations combine into a single rotation for each qubit, and for $\eta_{13}\ll\Gamma_{12}$, they only shift the effective rotation axis away from $\hat{x}$ slightly.  Similarly, the other additional couplings from all the disconnected peaks slightly shift the rotation axis for our simulated operations, though we found that these shifts are not large enough to prevent the demonstration of entanglement altogether.  More so, we stress that this alteration of our simulated operations stems from a non-ideal basis choice rather than any physics inherent to our proposal.

\begin{figure}[bht]
 \begin{center}
   \includegraphics[width=0.93\columnwidth]{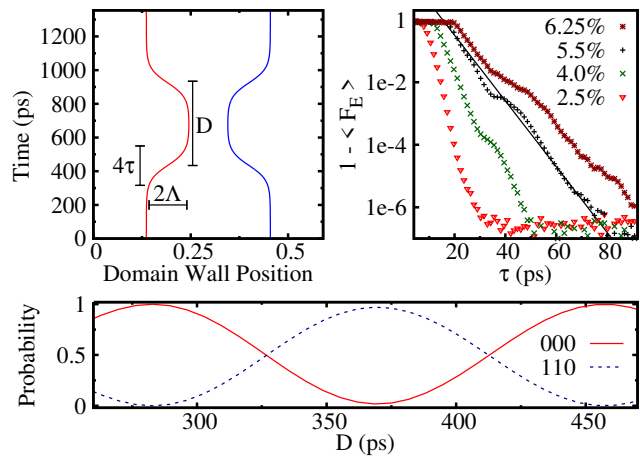}
 \caption{Top Left: Trajectories for the two left domain walls showing the amplitude, transition time, and duration for the $r_{12}^{x}(\pi/4)$ operation.  Top Right: Average infidelity of even states after performing $r_{12}^{x}(\pi/4)$, plotted against transition time for various amplitudes (labeled as percentages of the wire length $l$) showing exponential behavior in general agreement with the Landau-Zener formula until limited by our Runge-Kutta step-size.  The $\Lambda=0.055l$ data is fit with a line that scales as $\exp(-\beta\tau)$ with $\beta=240$ Ghz, reasonably close to the predicted value of $214$ GHz.  Bottom: Probabilities that the initial state $|000\rangle$ remains in $|000\rangle$ or transitions to $|110\rangle$ when acted on by $r_{12}^{x}$ with various duration times.  The simulated operation generally agrees with the expected rotation, but with a minimum probability of $\sim1\%$ for the $|000\rangle$ state due to a small shift in the rotation axis from the non-ideal basis choice in our simulation.}
   \label{fig:traj}
  \end{center}
\end{figure}

Once the chemical potential is tuned as described above, it can be varied dynamically to perform operations on the qubits.  The only operation needed to test entanglement inequalities are $r^{x}_{12}$ and $r^{x}_{23}$, which can be performed simultaneously by extending the outer topological regions towards the middle region.  Specifically the domain wall positions $x_{i}$ alternate back and forth according to the function
\begin{equation}
\pm \Lambda\left[\tanh\left(\frac{t}{\tau} \right)-\tanh\left(\frac{t-D}{\tau} \right)\right],
\end{equation}
which smoothly shifts the domain walls $2\Lambda$ over an approximate transition time of $4\tau$ for a duration $D$ between the center of the two transitions as shown in Fig.~\ref{fig:traj}.  

We are only concerned with the dynamics of the zero-energy states, which won't mix with the bulk states above the topological gap as long as the domain wall trajectories are adiabatic.  This constraint can be treated with the Landau-Zener condition:\cite{landau:1932, zener:1932} the rate the chemical potential changes must satisfy $\hbar |d\mu/dt|\ll2\pi\Delta_{\text{T}}^{2}$.  To test this in our simulation we find the probability that the basis states of Eqs.~$\eqref{basis}$ remain in the zero-energy subspace of the same total parity after evolution by using the following procedure. 

We assume the initial state $|\phi\rangle$ is in the set of even parity basis states given in Eqs.~\eqref{basis}, $S_{\text{E}}=\{|000\rangle,|011\rangle,|110\rangle,|101\rangle\}$, where $|000\rangle$ is defined as the state such that $\hat{d}_{n}|000\rangle=0$ for all $n$, including those corresponding to bulk states.  The zero-energy eigenstates are evolved using fourth-order Runge-Kutta in the eigenstate basis, possibly leaking into the bulk states if the transition time is too short, then acted on by the zero-energy eigenstate projector $\hat{P}_{0}$ to find the sub-matrix $\hat{U}_{0}=\hat{P}_{0}\hat{U}\hat{P}_{0}$ of the full evolution matrix $\hat{U}$.  Using $\hat{U}_{0}$, we find the time-evolved occupation operators in the Heisenberg picture, 
\begin{equation}
\hat{N}_{n}(t)=\hat{U}^{\dag}_{0}(t)\hat{d}^{\dag}_{n}(0)\hat{d}_{n}(0)\hat{U}_{0}(t),
\end{equation}
for $n=1,2,3$, which are bilinear combinations of the original $\hat{d}_{n}(0)$, including anomalous terms (e.g., $\hat{d}_{1}\hat{d}_{2}$) since the Hamiltonian contains superconductor pairing.  The three $\hat{N}_{n}(t)$ are then used to form the projector for each multi-particle state in our three-qubit basis, allowing us to calculate the probability that the corresponding state is occupied.  For example, the state $|110\rangle$ has the projector $\hat{N}_{1}\hat{N}_{2}(\hat{1}-\hat{N}_{3})$, which yields $0$ when acting on any other basis state.  Since the projector's eigenvalue for $|110\rangle$ is $1$, the expectation value is equal to the probability, and the probability that the initial state will be measured in the state $|110\rangle$ after time $t$ is
\begin{equation}
P_{110}(\phi)=\langle\phi|\hat{N}_{1}(t)\hat{N}_{2}(t)[\hat{1}-\hat{N}_{3}(t)]|\phi\rangle,
\end{equation}
which can be easily calculated for any $|\phi\rangle\in S_{\text{E}}$.  Similarly, the probabilities that $|\phi\rangle$ evolves into the other states in $S_{\text{E}}$ are found using their respective multi-particle projectors, which are summed to give the fidelity from $|\phi\rangle$ to $S_{\text{E}}$,
\begin{equation}
F_{\text{E}}(\phi)=P_{000}(\phi)+P_{011}(\phi)+P_{110}(\phi)+P_{101}(\phi).
\end{equation}
We calculate this for all $|\phi\rangle\in S_{\text{E}}$ after simulating the operation $r_{12}^{x}(\pi/4)$, and plot the average infidelity for even states, $1-\langle F_{\text{E}}\rangle$, versus the transition time for various amplitudes in Fig.~\ref{fig:traj}.  To compare our results with the Landau-Zener formula\cite{landau:1932, zener:1932} we use the maximum of the chemical potential rate
\begin{equation}
\frac{d\mu}{d t}=\frac{\partial\mu}{\partial x_{1}}\frac{\partial x_{1}}{\partial t},
\end{equation}
which occurs halfway through the transitions when $\mathrm{sech}^{2}(0)=1$, giving
\begin{equation}
\left(\frac{d\mu}{d t}\right)_{\text{max}}=\frac{\mu_{\text{T}}\Lambda}{\lambda\tau}.
\end{equation}
Thus $1-\langle F_{\text{E}}\rangle$ should scale as $\exp(-\beta\tau)$ with
\begin{equation}
\beta=\frac{2\pi\lambda\Delta_{\text{T}}^{2}}{\hbar \mu_{\text{T}}\Lambda}, \label{eq:fit}
\end{equation}
in general agreement with our data.  For example, the fitted line for $\Lambda=0.055l$ in the logarithmic plot in Fig.~\ref{fig:traj} has a slope that corresponds to $\beta_{\text{fit}}=240$ GHz, while the value predicted from Eq.~$\eqref{eq:fit}$, using $\Delta_{\text{T}}=0.36$ meV found in our simulation, is $\beta\simeq214$ GHz.  All the amplitudes fit the expected exponential behavior reasonably well until limited by our Runge-Kutta step-size, with the exception of small plateaus that appear at different transition times for different amplitudes.  This indicates that the coefficient for the average infidelity contains some amplitude-dependent factors, but these factors are insignificant compared to the exponential scaling and unimportant for our proposal.  We easily satisfy the adiabatic constraint by proceeding with our simulation using $\Lambda\sim0.055l$ and $\tau=75$ ps.

In order to ensure the $r^{x}_{12}$ operation is performed as expected, we find the probabilities for the basis states using the initial state $|000\rangle$ after the domain wall trajectory in Fig.~\ref{fig:traj} is simulated.  As anticipated, the probabilities $P_{000}(000)$ and $P_{110}(000)$ oscillate, with negligible probabilities (on the order of our step-size limit of $10^{-6}$) found in the states with incorrect total parity.  However, the operation doesn't complete a full bit-flip for the duration expected to correspond to a $\pi$ rotation, with $\sim1\%$ of the probability found in the $|011\rangle$ and $|101\rangle$ instead of $|110\rangle$.  This is consistent with a $1\%$ shift of the rotation axis away from $\hat{x}$ due to the non-localized eigenstate basis discussed above.  We find a similar shift in the axis for the operation $r^{x}_{23}$ when the right domain wall motion is simulated.

Using amplitudes $\Lambda_{L}=0.05575l$ and $\Lambda_{R}=0.055l$ for $r^{x}_{12}$ and $r^{x}_{23}$, respectively, the simulated rotations have a period of $\sim0.2$ns.  Since the operations are achieved by bringing together exponentially decaying peaks, the overlap-dependent coupling between topological regions (e.g., $\Gamma$ in the Majorana Hamiltonian) is exponentially sensitive to the trajectory amplitude.  Thus, longer rotation periods can be achieved by slightly decreasing the amplitude.  On the other hand, greater amplitudes give shorter periods, but can also risk fusing the adjacent Majorana bound states if increased too much, which begins to occur in our simulation near $\Lambda\sim0.065l$.  Thus, the typical operation time (including adiabatic transitions) for our parameters is on the order of $0.5$ ns.

\begin{figure}[bht]
 \begin{center}
   \includegraphics[width=0.9\columnwidth]{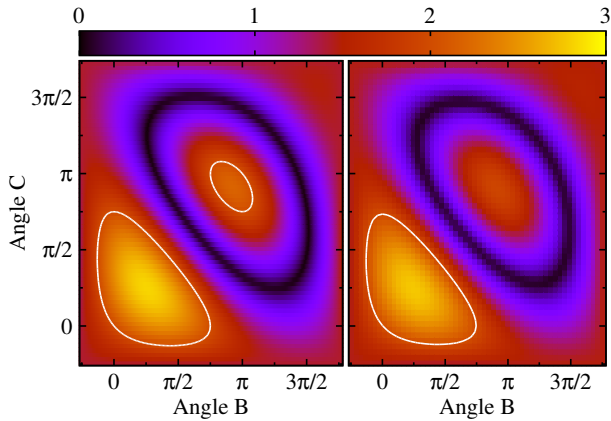}
 \caption{Left: Theoretical contour plot of the quantum mechanical prediction of the CHSH quantity for $|\Phi_\text{E}^{-}\rangle$ for the $A=\pi/4$ plane.  Local hidden variable theories are inconsistent with a CHSH quantity above 2, which occurs inside the white lines.  Right: Simulated contour plot showing general agreement except near the local maximum near $B=C=\pi$ due to a shift in the rotation axis for the operations.  Nonetheless, the global maximum at $B=C=\pi/4$ is still present and violates the CHSH inequality by approximately $40\%$. }
   \label{fig:chsh}
  \end{center}
\end{figure}

Finally, we test the CHSH inequality in our simulation by simultaneously performing the $r^{x}_{12}$ and $r^{x}_{23}$ rotations on the initial state $|\Phi_\text{E}^{-}\rangle$ and finding the probabilities for each basis state.  The CHSH quantity is a function of three relative angles, making it more difficult to visualize and compare to our simulation.  Instead we look at one plane involving the maximum violation, namely when $R_{2}=0$ and $L_{1}=\pi/4$ [i.e., $A=\pi/4$ in Eq.~\eqref{eq:chsh-angle}].  The theoretical prediction and simulation are plotted in Fig.~\ref{fig:chsh}, showing agreement except at the local maximum near $B=C=\pi$.  This difference is explained by the rotation axis shift discussed above, which we corroborated with additional simulations.  Nonetheless, the global maximum of $\sim2.8$ at $B\simeq C\simeq\pi/4$ is still present and there is a significant range of angles that violates the inequality.  Thus our simulation indicates that the more realistic semiconductor Hamiltonian is consistent with the simpler Majorana model and our proposal is feasible for demonstrating entanglement in a Majorana wire.

Before discussing experimental considerations, we introduce a simplification to the CHSH experiment that only requires projective measurement and the repeated use of two operations, namely $r^{x}_{12}(L)$ and $r^{x}_{23}(R)$ with specific values for $L$ and $R$.  Ideally $L=R=\pi/4$, but we leave them unspecified with the thought that the experiment could be attempted with angles that differ slightly from the ideal case.

The experiment begins by tuning the chemical potential to create three topological regions and measuring all of their occupation parities to project to one of the eight basis states.  For concreteness, we only consider the states that lead to the $|\Phi_\text{E}^{-}\rangle$ state, so the procedure only continues if the measurement of the middle parity matches the total parity [see Table~\ref{table:prep}].  Alternatively, the full experiment is carried out regardless of the measurement outcomes, but the cases when the parities do not match are disregarded.  Then the operations $r^{x}_{12}(L)$ and $r^{x}_{23}(R)$ are simultaneously performed twice before measuring the middle parity, proceeding only when this parity matches the initial result.  For the ideal angles $L=R=\pi/4$, this procedure prepares maximally entangled states.

\begin{figure}[bht]
 \begin{center}
   \includegraphics[width=\columnwidth]{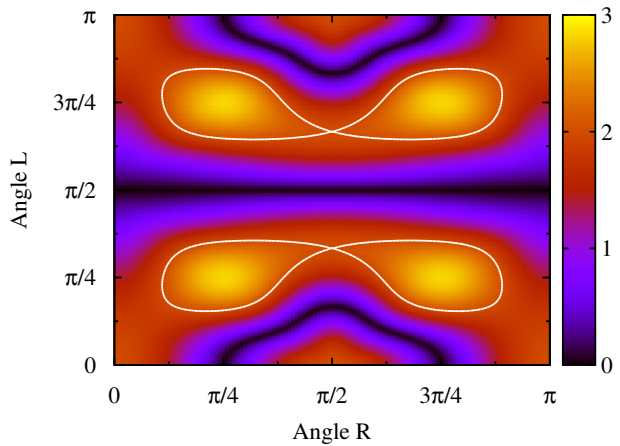}
 \caption{Contour plot of the quantum mechanical prediction of the CHSH quantity with $L_{1}=L$, $L_{2}=3L$, $R_{1}=0$ and $R_{2}=2R$.  Local hidden variable theories are inconsistent with a CHSH quantity above 2, which occurs inside the white lines.  The plot repeats with a period of $\pi$ for both $L$ and $R$. }
   \label{fig:chsh2}
  \end{center}
\end{figure}

This preparation is followed by one of the four rotation combinations in Eq.~\eqref{eq:chsh-orig}, with $L_{1}=L$, $L_{2}=3L$, $R_{1}=0$, and $R_{2}=2R$.  For example, the combination with $L_{2}$ and $R_{1}$ is performed by carrying out $r^{x}_{12}(L)$ three times but leaving the right domain walls stationary.  After one of the rotation combinations is performed, all three parities are measured and the results are recorded.  This is repeated several times for each combination to find the corresponding probabilities and calculate the CHSH quantity in Eq.~\eqref{eq:chsh-orig}.  The quantum mechanical prediction for the CHSH quantity using the above procedure is easily calculated (though not particularly illuminating in written form) and is plotted in Fig.~\ref{fig:chsh2}.  As expected, the CHSH quantity has a maximum at $L=R=\pi/4$, with a wide range of angles spanning from approximately $\pi/8$ to $3\pi/8$ confirming Bell's theorem.  Thus this procedure can be used for a broad range of angles, demonstrating entanglement in Majorana wires, even with limited accuracy in the tuning operations.

\section{Experimental Considerations}
\label{sec:experiment}
We now discuss some aspects of our proposal that may be significant for an experimental realization.  One of the first hurdles that must be overcome is the development of reliable projective measurement of the occupation.  Aside from directly probing the wire with point contacts, there have been several proposals for observing the presence of Majorana bounds states such as using the Aharanov-Casher effect,\cite{Hassler:2010} transmons,\cite{Hassler:2011} and Shapiro step doubling in the AC-Josephson effect.\cite{Rokhinson:nov2012}  Without committing to a particular readout scheme, we note that our proposal requires measurement of a single topological region during our procedure in order to project to a maximally entangled state.  

Another non-trivial experimental task is the fine-tuning of the chemical potential to minimize undesired Majorana peak overlaps.  The simplest way to mitigate these overlaps is to use a longer wire, which exponentially reduces the overlaps.  Our simulation indicates that a wire length on the order of $5-10 \mu$m is sufficient.  Alternatively, a setup that links together several shorter wires may also be possible if longer wires are experimentally unavailable.

The adiabatic constraint we found using the Landau-Zener formula is rather lenient, only requiring transition times on the order of $0.1$ns.  This is due to the generous topological gap of $\sim0.35$ meV that separated the zero-energy and bulk states, due to the relatively large proximity effect and $g$-factor.  In addition, the topological phase requires a relatively large spin-orbit coupling.  Our parameters are reasonable when compared to experiments,\cite{Mourik:Science2012, Deng:NanoLett2012, Das:2012, Finck:Mar2013, Rokhinson:nov2012} but the need for a robust topological gap should be considered when selecting materials, and further advances of the proximity effect and spin-orbit in relevant materials would be helpful.

The operation time for performing the ideal rotation angles can be found experimentally by reproducing the probability plot in Fig.~\ref{fig:traj}.  For example, the ideal $r^{x}_{12}(\pi/4)$ for the simplified CHSH experiment can be calibrated in the following way.  First, all three parities are measured to project to a basis state, then gates are gradually tuned to shift the chemical potential in the left non-topological region for an operation time $\sim0.5$ns, followed by a final measurement of all three parities.  We note that the operations can be achieved by moving a single domain wall if this is easier experimentally; we moved both the outer and middle domain walls to suppress overlaps in the topological regions, but this may be unnecessary in longer wires.  This procedure is repeated several times with the same operation time, tracking the percentage of times the state remains in the initially measured state.  This is repeated with several slightly different operation times, until the percentage is near $\cos^{2}(\pi/8)\simeq0.85$.  Once the rotations that correspond to $L=R=\pi/4$ are roughly calibrated, the simplified CHSH experiment can be carried out.

As noted by Rainis et. al.,\cite{Rainis:PRB2012} we must also consider the phenomena of quasi-particle poisoning in any system that uses superconductors to achieve the topological phase.  While a superconductor at $T=0$ (which we assumed in our simulation) will only form Cooper pairs, at finite temperatures less than $\sim160$mK a small, fixed population of quasi-particles seems to be present .\cite{deVisser:2011}  Quasi-particle poisoning occurs when a single quasi-particle tunnels between the superconductor and semiconductor, changing the total parity of the system and destroying the quantum information.  Thus, the measurement and operation times must be much shorter than the average time of quasi-particle tunneling, constraining the operation time in the opposite limit as the adiabatic condition.  Fortunately, estimates of the average time for quasi-particle tunneling in Majorana wire systems are on the order of $100$ns or greater\cite{Rainis:PRB2012}, leaving a wide window for the few operations needed in our proposal. 

Another non-trivial aspect inherent to the entanglement inequalities is the need to find probabilities rather than single measurements outcomes, requiring a high level of precision in the gate tuning.  While this may make it difficult to reproduce the exact predictions of quantum mechanics, the large violation of the CHSH inequality by $>40\%$, and the wide range of angles that violate hidden variable theories may still be sufficient for demonstrating entanglement.  

One way to circumvent the precision requirement is to perform the test proposed by Greenberger, Horne, Zeilinger,\cite{greenberger:90} and Mermin\cite{mermin:90} (GHZM). The GHZM experiment requires three logical qubits (thus four topological regions), but tests hidden variable theories with a single measurement, rather than involving probabilities and inequalities.  Indeed, there are many interesting experiments that demonstrate entanglement, such as quantum teleportation, that are possible with one additional qubit.   

This and other relevant systems are still a new and emerging area of research for theorists and experimentalists alike.  With that said, there are many recent developments which aren't reflected in our Hamiltonians.  For example, it seems that the experiments on the Majorana wire systems aren't strictly one-dimensional, and must be analyzed as multi-channel wires to explain some of the experimental findings.\cite{stanescu:2013}  Others note that phenomena like Andreev reflection, disorder, finite temperature and the Kondo effect may need to be further understood in these systems.\cite{stanescu:2013, Liu:Dec2012, Rainis:PRB13}  Despite these complications, we stress that almost any convincing manifestation of Majorana bound states must demonstrate entanglement, which will likely be easier than full braiding.   While we discussed the specific setup of Majorana wires, the general idea of using non-topological, proximity-induced operations to test the entanglement inequalities as a stepping stone to braiding operations, as well as other aspects like separating logical qubits with a parity qubit, may be applied to many systems that potentially support Majorana bound states, such as topological insulators.\cite{Fu:2008, Nilsson:Sep2008, Fu:Apr2009}

\section{Conclusion}
\label{sec:conclusion}
We have analyzed the motion of domain walls in a Majorana wire system with three separate topological regions using a simple Hamiltonian analogous to Kitaev's toy model.\cite{Kitaev:2001}  Using the occupation number parity of the Majorana bound states in each region, we defined a three qubit basis with two logical qubits and one qubit forfeited to total parity conservation.  In this basis, $x$-axis rotations are performed by extending the outer topological regions to isolate a single coupling between different topological regions.  While these rotations are not topologically protected from local perturbations, they can demonstrate entanglement by testing the Bell and CHSH inequalities.

With the simpler model as a guide, we simulated the domain wall motion using a more realistic semiconductor Hamiltonian.  Our results indicate that the topological regions can be well separated in wires of length $\sim5-10\mu$m with reasonable parameters compared to recent experiments.  Adiabatic changes in the chemical potential were simulated with operation times on the order of $0.5$ns, consistent with the Landau-Zener condition applied to excitations from zero-energy to the bulk.  Extending the topological regions results in the rotations predicted by the simpler model.  Finally, we simulated the CHSH experiment and found the expected inconsistency with hidden variable theories predicted by Bell's theorem, indicating that the simpler Majorana Hamiltonian approximates the Majorana wire system well.

We introduced a simplified version of the CHSH experiment that only requires projective measurement and repeated use of two $\pi/4$ rotations.  We found that a wide range of operation angles from $\pi/8$ to $3\pi/8$ violate the entanglement inequalities.  Thus a keyboard gating setup needs to be relatively precise, but moderate inaccuracy is tolerable.  We provided methods for calibrating the rotations and discussed potential hurdles for an experimental realization.  Our analysis and simulation indicate that there is a large window of operation times, spanning three orders of magnitude, that satisfy the adiabatic and quasi-particle poisoning constraints.  Hence our proposal is viable for demonstrating entanglement in Majorana wires if methods for projective measurement and precise gate tuning are available.  Experimental work on gate tuning is already required for braiding operations, and our proposal could serve as a useful benchmark towards that goal.  More so, the observation of entanglement would support current models of Majorana wires and provide a significant piece of evidence supporting the presence of Majorana bound states.

\section{Acknowledgements}  This work was
supported in part by the U.S. Army Research Office under Grant No.
W911NF-11-1-0027 (LPP, DD), and by the NSF under Grants DMR-0748925 (KS, DD) and
1018935 (LPP, DD), as well as the DARPA-QuEST program (DD, CYH and KS).  CYH would like to thank the funding support by the Packard Foundation.  AAK would like to thank the Central Facilities of
the Nebraska Center for Materials and Nanoscience supported by the Nebraska Research Initiative.  DD would like to thank Y. Barlas and A. De for helpful conversations.

\appendix
\section{General Preparation Procedure}
\label{sec:app2}

In Sec.~\ref{sec:inequalities} we discussed the Bell and CHSH inequalities with the state $|\Phi_{\text{E}}^{-}\rangle$.  Here we consider more general procedures for preparing any maximally entangled state and testing the inequalities.  Entanglement can be demonstrated for any initial condition with very simple alterations to our proposal, rather than discarding data for the incorrect initial state or measurement outcome.

The procedure shown in Fig.~\ref{fig:bell-prep} prepares one of the eight maximally entangled states,
\begin{eqnarray}
|\Phi_\text{E}^{\pm}\rangle=\frac{|000\rangle\pm|101\rangle}{\sqrt{2}},&\quad\quad&|\Psi_\text{E}^{\pm}\rangle=\frac{|011\rangle\pm|110\rangle}{\sqrt{2}},\quad\quad\\
|\Phi_\text{O}^{\pm}\rangle=\frac{|010\rangle\pm|111\rangle}{\sqrt{2}},&\quad\quad&|\Psi_\text{O}^{\pm}\rangle=\frac{|001\rangle\pm|100\rangle}{\sqrt{2}},\quad\quad
\end{eqnarray}
by measuring all three parities to project to one basis state from Eqs.~\eqref{basis}, performing the operations $r^{x}_{12}(\pi/2)$ and $r^{x}_{23}(\pi/2)$, then measuring the middle parity.  The state that is prepared depends on the overall parity and middle parity measurements, as shown in Table~\ref{table:prep}.  Note that the results for the even and odd total parity are equivalent upon the exchange $0\leftrightarrow 1$ for the middle parity qubit.

\begin{table}[ht]
\caption{\label{table:prep} Maximally entangled state prepared for various total parity and middle parity measurements.  The even and odd total parities give the same results if we exchange $0\leftrightarrow 1$ for the middle parity. }
\begin{ruledtabular}
\begin{tabular}{ | l | c | c | c | c | c | c | c | c | }
\hline
Total Parity & \multicolumn{4}{c |}{Even} & \multicolumn{4}{c |}{Odd} \\
\hline
Initial Middle Parity & \multicolumn{2}{c |}{0} & \multicolumn{2}{c |}{1} & \multicolumn{2}{c |}{1} & \multicolumn{2}{c|}{0}  \\
\hline
Final Middle Parity & 0 & 1 & 0 & 1 & 1 & 0 & 1 & 0 \\
\hline 
Resulting State & $\Phi_{\text{E}}^{-}$ & $\Psi_{\text{E}}^{+}$ & $\Phi_{\text{E}}^{+}$ & $\Psi_{\text{E}}^{-}$ & $\Phi_{\text{O}}^{-}$ & $\Psi_{\text{O}}^{+}$ & $\Phi_{\text{O}}^{+}$ & $\Psi_{\text{O}}^{-}$ \\
\hline 
\end{tabular}
\end{ruledtabular}
\end{table}

Any of the maximally entangled states can be used to demonstrate the violation of the Bell and CHSH inequalities, but with different rotation angles.  For example, quantum mechanics predicts that $\langle L,R\rangle$, the expectation value of the combined parity of the left and right qubits after being rotated by angles $L$ and $R$, respectively, for $\Phi_{\text{E}}^{+}$ is $\cos(L+R)$ rather than $\cos(L-R)$ for $\Phi_{\text{E}}^{-}$.  Clearly the CHSH quantity  in Eq.~\eqref{eq:chsh-orig} is the same except with $R\rightarrow-R$, which can be returned to the case in Sec.~\ref{sec:inequalities} by substituting $\{L_{1}, L_{2}, R_{1}, R_{2}\}\rightarrow\{L_{1}, L_{2}, -R_{1}, -R_{2}\}$.  The relevant probabilities and angle transformations for the Bell and CHSH inequalities are listed in Table~\ref{table:probs} for each maximally entangled state.  These changes can be accounted for by designing the experiment to perform different rotations depending on the middle parity measurement outcomes found during the preparation of the maximally entangled states.   

\begin{table}[h]
\caption{\label{table:probs} Probabilities and expectation values predicted by quantum mechanics for the  various maximally entangled states.  The set of angles that corresponds to the case in the body of the paper for the CHSH violation is given as well.  The results are the same for even and odd parity, so we suppress the corresponding subscript. }
\begin{ruledtabular}
\begin{tabular}{ | c | c | c | c | }
\hline
State & $P_{=}(L,R)$ & $\langle L, R \rangle$ & CHSH Angles \\
\hline
$\Phi^{-}$ & $\cos^{2}\left(\frac{L-R}{2}\right)$ & $\cos(L-R)$ & $L_{1}, L_{2}, R_{1}, R_{2}$\\
\hline
$\Phi^{+}$ & $\cos^{2}\left(\frac{L+R}{2}\right)$ &$\cos(L+R)$  & $L_{1}, L_{2}, -R_{1}, -R_{2}$ \\
\hline 
$\Psi^{-}$ & $\sin^{2}\left(\frac{L-R}{2}\right)$ & $-\cos(L-R)$ & $L_{1}, L_{2}, R_{1}, R_{2}$ \\
\hline
$\Psi^{+}$ & $\sin^{2}\left(\frac{L+R}{2}\right)$ & $-\cos(L+R)$  & $L_{1}, L_{2}, -R_{1}, -R_{2}$ \\
\hline 
\end{tabular}
\end{ruledtabular}
\end{table}

\section{Entanglement Inequalities Review}
\label{sec:app1}
In this appendix we briefly review Bell's theorem and the entanglement inequalities.  We only cover the basic aspects needed for our proposal; we refer the interested reader to the numerous works\cite{preskill:lec, mermin:90, selleri:1990} on the topic for a more comprehensive review.  

According to quantum mechanics, some multi-particle systems can only be described as a single entangled state.  For example, the maximally entangled two-qubit systems cannot be written as separable states; measuring one of the qubits automatically determines the outcome of the other.  We derive a simplified version of Bell's inequality by considering an experiment that separately measures each qubit of the state $|\Phi^{-}\rangle=(|00\rangle-|11\rangle)/\sqrt{2}$ with one of three different methods, denoted $a$, $b$, and $c$.  When both qubits are measured using the same method, the results are always the same.  However, when the two qubits are measured using different methods, the results are completely uncorrelated.  Thus the possible results for one qubit measurement depend on what method is used for the other qubit, even if the measurement events are well separated spatially (i.e., space-like).

Einstein, Podolsky, and Rosen (EPR) famously objected to this type of non-local behavior,\cite{einstein:35} citing it as motivation for a more complete theory that removes the probabilistic nature of quantum mechanics by introducing ``hidden variables''.  Hidden variable theories predict, with full certainty, the outcomes of different measurement methods on a single qubit, even though only one measurement at a time is possible.  Bell's theorem states that any local hidden variable theory makes predictions that are inconsistent with quantum mechanics.\cite{bell:1964}  Thus any experiment that agrees with quantum mechanics rather than hidden variable theories, implies that the qubits in the system are entangled.

To see where the two theories are inconsistent, we discuss the interpretation of $|\Phi^{-}\rangle$ in hidden variable theories.  Instead of a pure state, it is viewed as a classical ensemble of states, prepared with different hidden variables.  If an experimentalist could measure a single preparation with all three methods at once, the two qubits' results would match for each method.  In this view the two qubits only seem uncorrelated when using different methods, but are actually correlated regardless of the method chosen.  Thus the possible results of one qubit measurement don't depend on the method chosen for the other.

While this interpretation avoids non-local behavior, it replaces a superposition of states with a classical ensemble.  Thus any single preparation in the ensemble must be either $0$ or $1$.  Since there are three measurement methods, but only two possible outcomes, the pigeonhole principle states that at least two of the  methods must give matching results.  By defining $P_{=}(a,b)$ as the probability that the results match when one qubit is measured with $a$ and the other is measured with $b$, this statement can be written
\begin{equation}
P_{=}(a,b)+P_{=}(b,c)+P_{=}(a,c)\geq1,
\end{equation}
which is one version of Bell's inequality.  Meanwhile, quantum mechanics predicts that this inequality is invalid for certain measurement methods, which demonstrates Bell's theorem.

While this inequality can be tested experimentally in principle, it requires method $b$ to be tested for both qubits, which would be difficult to accomplish exactly in our proposal.  Instead, we look at the case where the left qubit of $|\Phi^{-}\rangle$ is measured with either method $L_{1}$ or $L_{2}$, while the right qubit is measured using either method $R_{1}$ or $R_{2}$.  Without superposition, each hidden variable preparation of the left qubit must have either $L_{1}=0$ or $L_{1}=1$, by which we mean that measuring the left qubit with method $L_{1}$ would yield $0$ or $1$, respectively.  It is simpler to derive the inequality by considering the parity of these quantities so we use $1$ and $-1$ for even and odd parity, respectively, for the remainder of this appendix.  Thus, each preparation must have $L_{1}$, $L_{2}$, $R_{1}$, and $R_{2}$ as either $1$ or $-1$ according to the hidden variable interpretation.

Consider the quantities $L_{1}+L_{2}$ and $L_{1}-L_{2}$; either $L_{1}+L_{2}=\pm2$ and $L_{1}-L_{2}=0$, or $L_{1}+L_{2}=0$ and $L_{1}-L_{2}=\pm2$.  This implies that the quantity
\begin{equation}
|(L_{1}+L_{2})R_{1}+(L_{1}-L_{2})R_{2}|=2 \label{eq:chsh1}
\end{equation}
for each preparation since one of the terms vanishes in either case.  

If an experimentalist could measure the qubits with more than one method at a time, this prediction could be tested directly.  Instead, we must extract a statistical prediction that only requires a single measurement of each qubit for any given preparation.  With that in mind, we note that the expectation value of any constant is simply that constant, and any variable $X$ satisfies $|\langle X\rangle|\leq\langle |X|\rangle$ for any probability distribution.  Applying these arguments to  Eq.~$\eqref{eq:chsh1}$, we get the eponymous inequality first derived by Clauser, Horne, Shimony, and Holt\cite{clauser:69}
\begin{equation}
\left|\langle L_{1},R_{1}\rangle+\langle L_{2},R_{1} \rangle+\langle L_{1},R_{2} \rangle-\langle L_{2},R_{2} \rangle\right|\leq2,
\end{equation}
where $\langle L,R\rangle=P_{=}(L,R)-P_{\neq}(L,R)$ is the expectation value for the combined parity of the left and right qubits when measured with methods $L$ and $R$, respectively.  Since each term only involves one measurement per qubit, it is possible to predict the left side of the inequality with quantum mechanics.  For several measurement method combinations, the predictions are inconsistent with the local hidden variably theories.  Thus any experiment that violates the CHSH inequality negates the local hidden variable theories, demonstrating entanglement in the system.

%\bibliography{Mwire}

%merlin.mbs 2010-03-15 4.21a (PWD, AO, DPC)

\end{document}